\documentclass[twocolumn,amsmath,amssymb,prl,10pt,nofootinbib,superscriptaddress]{revtex4}
\usepackage{ graphicx, float,amsmath, amsmath, amssymb}
\usepackage[export]{adjustbox}

\def\be{\begin{equation}}
\def\ee{\end{equation}}
\def\bea{\begin{eqnarray}}
\def\eea{\end{eqnarray}}
\def\bse{\begin{subequations}}
\def\ese{\end{subequations}}

\usepackage[breaklinks, colorlinks, citecolor=blue]{hyperref}
\usepackage[labelsep=period]{caption}
\usepackage[normalem]{ulem}
\usepackage{soul}
\usepackage{minitoc}

\usepackage{bm}

\begin{document}
\title{Quantum fields in the background spacetime of a polymeric loop black hole}

\author{Flora Moulin}%
\affiliation{%
Laboratoire de Physique Subatomique et de Cosmologie, Universit\'e Grenoble-Alpes, CNRS/IN2P3\\
53, avenue des Martyrs, 38026 Grenoble cedex, France
}

\author{Killian Martineau}%
\affiliation{%
Laboratoire de Physique Subatomique et de Cosmologie, Universit\'e Grenoble-Alpes, CNRS/IN2P3\\
53, avenue des Martyrs, 38026 Grenoble cedex, France
}

\author{Julien Grain}%
\affiliation{%
Institut d'astrophysique spatiale, Universit\'e Paris-Sud, CNRS \\
B\^atiments 120 \`a 121, Universit\'e Paris Sud, 91405 ORSAY, France
}%

\author{Aur\'elien Barrau}%
\affiliation{%
Laboratoire de Physique Subatomique et de Cosmologie, Universit\'e Grenoble-Alpes, CNRS/IN2P3\\
53, avenue des Martyrs, 38026 Grenoble cedex, France
}

\date{\today}
\begin{abstract} 
The description of black holes in loop quantum gravity is a hard and tricky task. In this article, we focus on a minisuperspace approach based on a polymerization procedure. We consider the resulting effective metric and study the propagation of quantum fields in this background. The cross sections for scalar particles and fermions are explicitly calculated. The Teukolsky-Chandrasekhar procedure used to derived the fermionic radial equation of motion for usual spacetimes is entirely generalized to a much larger class. The resulting radial equation can be used in quite a lot of other contexts.
\end{abstract}
\maketitle

\section{Introduction}

Loop quantum gravity (LQG) is a mature framework which is mathematically consistent and can be approached by several complementary paths, from canonical quantization to spinfoams (see, {\it e.g.}, \cite{Rovelli:2004tv,Thiemann:2007zz,Rovelli:2014ssa,Ashtekar:2017yom} and references therein). The ideas of the theory have been successfully applied to the Universe, leading to the loop quantum cosmology (LQC) paradigm (see, {\it e.g.}, the reviews \cite{Bojowald:2008zzb,lqc9,Barrau:2013ula,lqc1,Agullo:2013dla,Wilson-Ewing:2016yan,Agullo:2016tjh,Barrau:2016nwy}, and references therein) and to black holes (BHs) (see, {\it e.g.}, the reviews \cite{Ashtekar:2000eq,Gambini:2013exa,G.:2015sda,Olmedo:2016ddn,Perez:2017cmj}, and references therein).\\

In this article, we focus on the BH issue and consider the propagation of quantum fields. There are many different attempts to deal with BHs in LQG and to describe their dynamics. In this study, we use an effective corrected metric derived in \cite{Alesci:2011wn}. This spacetime structure is in no way a final word on the question of the exterior background of an LQG BH. It relies on heavy hypothesis that should be questioned. But it constitutes an interesting phenomenological framework to investigate the questions of cross-sections and greybody factors in an effective quantum gravity-corrected background.  Within this spacetime, we investigate in details the scattering of quantum fields. We first draw the general picture used to model BHs in this framework. Then we explain how cross sections are calculated and their meaning. We turn to the explicit computation for scalar particles. Finally, we derive the propagation equation for fermions. Conclusions and perspectives are outlined.

\section{Black holes in loop gravity}

BHs are fascinating objects that have been intensively investigated in the framework of loop quantum gravity \cite{Ashtekar:2000eq,Gambini:2013exa,G.:2015sda,Olmedo:2016ddn,Perez:2017cmj}. To give just one example, the Bekenstein-Hawking entropy is now correctly recovered, although different ways to compute it are still considered (see, {\it e.g.} \cite{Perez:2017cmj}). In microcanonical calculations taking into account only the quantum geometrical degrees of freedom \cite{Rovelli:1996dv} this requires a specific fixing of the Barbero-Immirzi parameter, depending on the details of the state counting \cite{Agullo:2009eq}. This is not anymore the case in recent holographic models \cite{Ghosh:2013iwa,Frodden:2012dq,Achour:2014eqa,Asin:2014gta,Achour:2015xga}.\\

In this study, we use the metric obtained in \cite{Alesci:2011wn}, building on \cite{Modesto:2008im}. This framework was precisely set-up to investigate the creation of BHs and their subsequent Hawking evaporation. This question is intimately related to the information paradox which is itself closely linked to the singularity resolution. An interesting approach consists in using the 4-dimensional static model derived in \cite{Modesto:2008im} and to make it dynamical. This allows one to reproduce the Hawking calculation of particle creation in a classical BH background and to demonstrate that the whole process is unitary. The spirit of the framework in the line of the long history of ``non-singular" BHs (see, {\it e.g.}, \cite{Frolov:1989pf,Balbinot:1991qda,Hayward:2005gi,Spallucci:2009zz}, and references therein).\\

In canonical LQG, the basic variables are the holonomy of the Asktekar connexion and the flux of the densitized triads. In the covariant formulation, space is described by a spin network whose edges are labelled by irreductibles representations of SU(2) and nodes are intertwiners \cite{lqg3}. Intuitively, the edges carry quanta of area and the vertices carry elementary volumes. One of the most important result of LQG is that the area is quantized according to:
\begin{equation}
A(j)=8\pi\gamma l_P^2\sqrt{j(j+1)},
\label{area}
\end{equation}
where $\gamma$ is the Barbero-Immirzi parameter, $l_P$ is the Planck length and $j$ is a half-integer. In \cite{Alesci:2011wn}, several hypotheses were made to describe LQG BHs beginning, as expected, by spherical symmetry which is used to reduce the number of variables. In addition, instead of all {\it a priori} possible closed graphs, a regular lattice with edges of lengths $\delta_b$  and $\delta_c$ has been chosen. Details on the structure of lattices possibly used can be found in \cite{Alesci:2015nja}. The resulting dynamical solution inside the horizon was then analytically continued to the region outside the horizon, showing that it is possible to reduce the two unknown parameters by requiring that the minimum area in the solution is equal to minimum area of LQG (exactly as done in LQC). The remaining free parameter $\delta_b$ will now be called $\delta$ and referred to as the ``polymeric parameter". Together with $A_{min}=A(1/2)$, it determines how ``different" from the usual general relativity (GR) solution the considered BH is. \\

\begin{figure}
\centering
    \includegraphics[width=.9\linewidth]{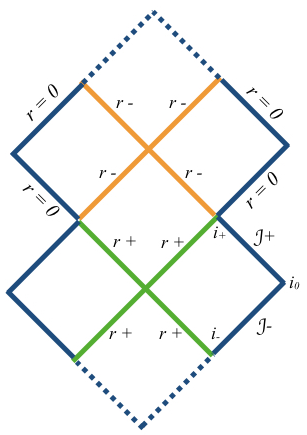}  
 \caption{Penrose diagram for the metric considered in this study. The horizons are denoted as $r_+$ and $r_-$.}
      \label{penrose}

\end{figure}

In practice, the procedure consists in first defining the Hamiltonian constraint by the use of holonomies along the considered fixed graph. It is important to underline  that the influence of the choice of a specific graph  has not been studied in details and this should be considered as a weakness of the considered approach. Both the diffeomorphism and Gauss constraints are identically vanishing: the first one is zero because of homogeneity and the second one is zero because the spacetime is of the Kantowski-Sachs form. The Hamiltonian constraint is solved after replacing the connection by the holonomy. Finally, the solution is expanded to the full spacetime, leading to the effective LQG-corrected geodesically complete Schwarzschild metric:
\begin{eqnarray}
&& ds^2 =  G(r) dt^2 - \frac{dr^2}{F(r)} - H(r) d\Omega^2~, \nonumber \\
&& G(r) = \frac{(r-r_+)(r-r_-)(r+ r_{x})^2}{r^4 +a_o^2}~ , \nonumber \\
&& F(r) = \frac{(r-r_+)(r-r_-) r^4}{(r+ r_{x})^2 (r^4 +a_o^2)} ~, \nonumber \\
&& H(r) = r^2 + \frac{a_o^2}{r^2}~,
\label{g}
\end{eqnarray}
where $d \Omega^2 = d \theta^2 + \sin^2 \theta d \phi^2$, $r_+ = 2m$ and $r_-= 2 m P^2$ are the two horizons (being respectively future and past horizons for observers in the two asymptotically flat regions of the associated causal diagram), and $r_x = \sqrt{r_+ r_-} = 2mP$, $P$ being the polymeric function defined by $P = (\sqrt{1+\epsilon^2} -1)/(\sqrt{1+\epsilon^2} +1)$, with $\epsilon=\gamma\delta$, and the area parameter $a_o$ is given by $a_0=A_{min}/8 \pi$. In principle $\epsilon$ is not bounded but the approach is rigorous only when $\epsilon\ll 1$ (at this state no phenomenological bound has been derived on $\epsilon$). The parameter $m$ in the solution is related to the ADM mass $M$ by $M = m (1+P)^2$ (inferred from what is observed at asymptotic infinity). This metric should be considered as a ``toy model" and not taken as a final statement about the spacetime structure around an LGQ BH. It is however very convenient and meaningful for first phenomenological investigations. The associated Penrose digram is given in Fig \ref{penrose}. From now, we use only Planck units. \\

Let us discuss a bit more this solution. The considered spacetime is a particular example of a Kantowski-Sachs spacetime. In the construction, the interior of a spherically symmetric BH is treated as homogeneous, but not explicitly as isotropic. As usual, the connection is replaced by the holonomy in the Hamiltonian constraint and the equation of motion are solved, together with the Hamiltonian constraint.  The outcome is an exact solution of a minisuperspace model valid inside the event horizon  \cite{Modesto:2008im}. Finally, the solution is analytically extended to the whole spacetime. In other words, the metric was assumed to be valid everywhere and it was explicitly proven with a coordinate transformation that the singularities at the two horizons (event horizon and Cauchy internal horizon) were just coordinate singularities. The resulting metric has a simple, geodesically complete, analytic form in the whole spacetime. The weaknesses are the following. First, the metric cannot be considered a rigorous ``full LQG" solution, although it captures some features of LQG as the minimum area and the use of holonomies. Second, this metric builds on the initial version of LQC. In the future it would be interesting to replace the polymeric parameter by a rescaled one, in the same sense than the $\mu_0$ scheme in LQC has been replaced by the $\bar{\mu}$ one (see \cite{lqc9}). Finally, it is assumed that matter couples minimally to the effective metric.\\

It should me underlined that the model considered in this article is far from being the only possible one within the LQG framework. It is somehow ``unusual" in the sense that it might lead to possible large quantum gravity effects outside the horizon. Although not something fully exotic (this possibility is {\it e.g.} advocated on a different grounding in \cite{Haggard:2016ibp}), it is fair to say that this is not a generic prediction. It is however the specific case where quantum gravity might have an impact on observations and this is why we focus here on this specific setting which is anyway quite well justified in its physical motivations. 

\section{Cross section for evaporating black holes}

The Hawking evaporation \cite{Hawkin} (as a specific case of the Unruh effect \cite{Unruh:1976db}) is one of the most important aspects of BH physics. Although it can be described as semi-classical process in the ``large mass" regime, it requires a quantum gravity treatment near the endpoint. Several attempts to describe it in the framework of LQG were made \cite{Ashtekar:2005cj,Barrau:2011md,Gambini:2013nea}. In this study we focus on another aspect. Basically, the ``naive" Hawking spectrum is described by a blackbody law, in agreement with the Unruh effect which predicts that an accelerated observer sees a bath of thermal particles with temperature $T=a/(2\pi)$. In the case of black holes, the temperature is $T_H=1/(8 \pi M)$: the lighter the BH, the highest its temperature, which makes the whole process very explosive in the last stages (a BH with a mass above the mass of the Moon has a temperature smaller that the one of the cosmological microwave background). However, the real spectrum is slightly more complicated as the emitted particles have to cross a potential barrier before escaping to infinity. This induces a modification, captured by the cross section $\sigma$, to the pure blackbody spectrum which is known to encode quite a lot of information on the gravitational theory or spacetime structure considered. The spectrum reads as:

\be
\frac{dN}{dt}= \frac{1}{e^{\frac{\omega}{T_H}} \pm 1} \sigma(M,s,\omega) \frac{d^3k}{(2 \pi )^3}, 
\ee

\noindent with $M$ the BH mass, $s$ the particle spin, $\omega$ its energy and $k$ its momentum.\\ 

Cross sections have already been calculated for many metrics, beginning by the pioneering works on  Schwarzschild, Kerr, and Reisner-Nordstrom BHs in the case of scalar, fermion and vector fields \cite{DNPagescalar, DNPagedirac, DNPagevector}. They have also been investigated for extra-dimensional Schwarzschild-de-Sitter black hole \cite{Kanti:2005ja}, for lovelock gravity \cite{Grain:2005my}, for tachyonic fields \cite{Gursel:2018bts}, for scalar fields in an Einstein-Maxwell background \cite{Panotopoulos:2018pvu}, for $f(R)$ gravity minimally coupled to a cloud of strings in $2+1$ dimensions \cite{Ovgun:2018gwt}, for Einstein-Gauss-Bonnet–de Sitter black holes \cite{Zhang:2017yfu}, for black strings \cite{Ahmed:2017edq}, for Einstein-Born-Infeld dilaton spacetimes \cite{Panotopoulos:2017yoe}, for dRGT massive gravity \cite{Boonserm:2017qcq}, for Reissner–Nordström–de Sitter black holes \cite{Ahmed:2016lou}, for extra-dimensional Kerr black holes \cite{Jorge:2014kra}, for Myers-Perry black holes \cite{Boonserm:2014fja}, for dilatonic black holes \cite{Abedi:2013xua}, for rotating charged Goedel black holes \cite{Li:2009zzf}, to cite only a few remarkable results. In each case the cross section captures some specific and non-trivial characteristics of the considered spacetime.
In this article, we calculate the cross sections for a so-called loop BH (LBH), as described by the metric (\ref{g}), which is static and spherical symmetric. Given those spacetime symmetries, and according to the optical theorem, the cross section reads

\be 
\sigma(M,s,\omega) = \sum_{l=0}^{ \infty} \frac{(2j+1) \pi}{\omega ^2} |A_{l,s} |^2,
\label{sigma}
\ee

\noindent where $A_{l,s} $ is the transmission coefficient of the angular momentum mode $l$, and $j=l+s$ is the total angular momentum.

\section{Massless Scalar Field}

The dynamics of a massless scalar field minimally coupled to the gravitational field is described by the generalized Klein-Gordon equation:

\be
\frac{1}{\sqrt{-g}}\partial_{\mu} (g^{\mu \nu }\sqrt{-g}\partial_{\nu} \Phi)=0,
\label{kg}
\ee

\noindent where \textbf{ $\Phi \equiv \Phi(t,r, \theta, \phi)$}. Since we work within a static and spherically symmetric setting, the scalar field can be written as: 

\be
\Phi(r, \theta, \phi ,t)=R(r)S(\theta)e^{i(\omega t+m' \phi)},
\ee

\noindent where $w$ is the frequency and $m'$ is an integer. When inserting this ansatz in the Klein-Gordon equation (\ref{kg}) with the metric (\ref{g}), the radial equation reads

\be
\frac{\sqrt{GF}}{H} \frac{\partial }{\partial  r} \left( H\sqrt{GF}\frac{\partial  R(r)}{\partial  r}\right)+\left(\omega ^2 - \frac{G}{H} l(l+1) \right) R(r)=0,
\label{radialr}
\ee

\noindent with $l$ the orbital quantum number. This result uses the squared angular momentum operator $L^2=- \left[ \frac{1}{sin^2 \theta } \frac{\partial ^2}{\partial \phi ^2} +  \frac{1}{sin \theta } \frac{\partial }{\partial \theta } \left( sin \theta \frac{\partial }{\partial \theta} \right) \right] $,  whose eigenvalues are $l(l+1)$. \\

As usually done to study this kind of problems, we introduce the tortoise coordinate. Focusing on the two non-trivial coordinates, the metrics (\ref{g}) reduces to

\be
ds^2  =  -G(r)dt^2+\frac{dr^2}{F(r)},
\ee

\noindent and the null geodesics are given by $ds^2=0$, that is $dt^2=\frac{dr^2}{GF} \equiv dr^{* 2}$ with $r^*$ the tortoise coordinate. This new coordinate tends to $- \infty$ when $r$ tends to $r_+$. By introducing a new radial field $\Psi (r) \equiv \sqrt{H} R(r)$ and writing Eq. (\ref{radialr}) with respect to $r^*$, we obtain:

\be
\left( \frac{\partial ^2 }{\partial  r^{* 2}} +\omega ^2 - V(r^*) \right) \Psi (r) = 0,
\ee
\be
V(r) = \frac{G}{H} l(l+1)+\frac{1}{2}\sqrt{\frac{GF}{H}}\frac{\partial}{\partial r}\left(\sqrt{\frac{GF}{H}}\frac{\partial H}{\partial r} \right).
\ee

\noindent The potential $V(r)$ vanishes at the horizon $r_+$ and at spatial infinity. \\

At the horizon $r_+$, $\sqrt{H}$ tends to the constant $\sqrt{H(r_p)}$ and the radial part of the wavefunction $R$ is a plane wave with respect to the tortoise coordinate: 

\be
R (r^*)= A^h_{in}e^{i \omega r^*}+A^h_{out}e^{-i \omega r^*},
\label{horizon}
\ee

\noindent with $A^h_{in}$ (respectively $A^h_{out}$) the probability amplitude for the incoming modes (resp. outgoing modes) at the horizon. For convenience, we choose the absorption point of view. With this convention, there are incoming and outgoing modes infinitely far from the BH and only incoming ones at the horizon. We therefore impose $A^h_{out}=0$. \\

Infinitely far away from the horizon, $\sqrt{H}$ tends to $r$ and the radial wavefunction is a spherical wave with respect to the coordinate r:

\be
R(r)= \frac{A^{\infty}_{in}}{r}e^{i \omega r}+\frac{A^{\infty}_{out}}{r}e^{-i \omega r}.
\label{infini}
\ee

For a scalar particle, the transmission amplitude for the mode $l$ is given by:

\be
|A_l |^2=r_+^{2} \left| \frac{A^{h}_{in}}{A^{\infty}_{in}} \right| ^2 = 1-\left| \frac{A^{\infty}_{out}}{A^{\infty}_{in}} \right| ^2.
\label{coe}
\ee

The calculation of the cross section relies on the following steps. For each quantum number $l$, we solve the radial equation (\ref{radialr}) so as to determine the transmission coefficients $A^{\infty}$. Numerical computations must be performed from the horizon (where the radial wavefunction takes the form of Eq. (\ref{horizon})) until infinity (where the radial wave function takes the form of Eq. (\ref{infini})). In practice, the numerical solving begins at $r_{ini}= r_+ + 10^{-3}r_+ $ and stops sufficiently far at $r_{end}  \approx 300/ \omega$ which can be considered as infinity at the chosen accuracy. 

We decompose the radial wavefunction $R(r)$ into its real part $U(r)$ and its imaginary part $V(r)$. At $r_{ini} \approx r_+ $, the normalization condition $R(r_{ini})=1$ ensures that there are only incoming modes and $\frac{d R(r_{ini})}{dr}=\frac{i \omega}{\sqrt{G(r_{ini}) F(r_{ini})}} $. Technical details are given in Appendix A. 

\begin{figure}
\centering
    \includegraphics[width=.9\linewidth]{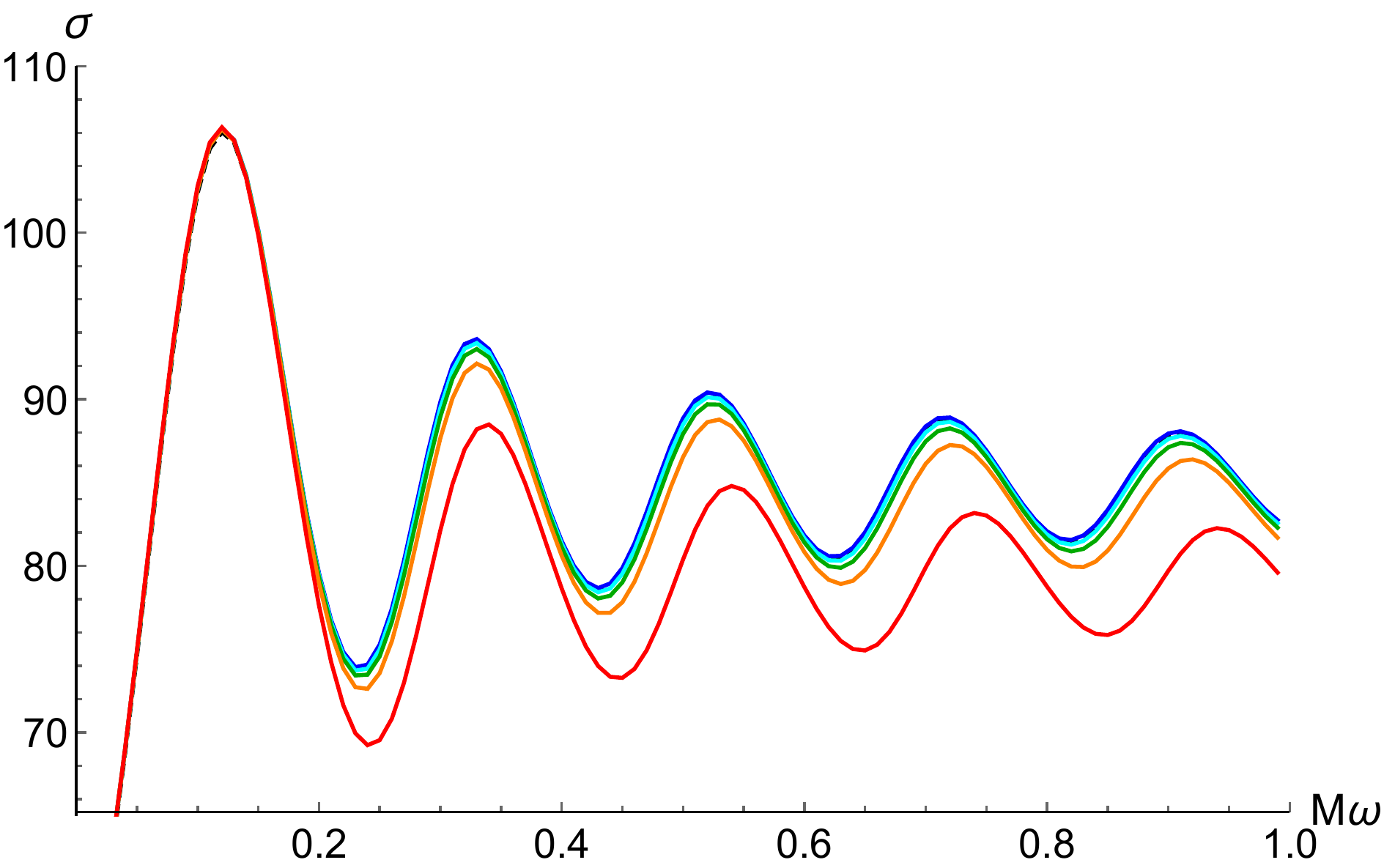}  
 \caption{Emission cross section for a scalar field with energy $\omega$ in the background spacetime of a LBH of mass $M$ for different values of $\epsilon$ ($\epsilon=\gamma\delta$ measures the ``quantumness" of spacetime). From bottom to top: $\epsilon=10^{ \{-0.3,-0.6,-0.8,-1,-3\}}$. The blue line, corresponding to $\epsilon=10^{-3}$ is superposed with the cross section for a Schwarzschild BH.}
      \label{CSection}

\end{figure}

The radial equation is solved with a fifth order Runge Kutta method until $r_{end}$. The  radial wavefunction is fitted with the function given by Eq. (\ref{infini}) so as to obtain the coefficients $A^{\infty}_{out}$ and $A^{\infty}_{in}$. Then the $|A_l |^2$ can be obtained from Eq. (\ref{coe}). The bigger the $l$, the smaller the $|A_l |^2$ and numerical investigations have shown that stopping at  $l=10 $ is sufficient. Finally, equation (\ref{sigma}) is used to evaluate the cross section. The results are presented in figure \ref{CSection}. \\

The cross section does decrease when $\epsilon$ increases. One can also notice a slight energy shift of the pseudo-periodic oscillations toward a lower frequency (in $M\omega$) when  $\epsilon$ increases. When $\epsilon < 10^{-0.8}$, it is hard to distinguish between the solutions. As far as phenomenology is concerned, it seems that taking into account the quantum corrections does not influence substantially the cross section of a scalar field for reasonable values of $\epsilon$ (that is $\epsilon
\ll 1$). The main trend is however clear.

\section{Spin $\frac{1}{2}$ field }


For spacetimes such that $ds^2=f(r)dt^2-f^{-1}(r)dr^2-r^2 d \Omega ^2$,  the radial equation is given by the Teukolsky master equation \cite{Teukolsky:1973ha}. The metric given by Eq. (\ref{g}), without any specified expressions for $G(r)$, $F(r)$ and $H(r)$, is however more general and basically includes all the static and spherical spacetimes. To the best of our knowledge, the fermionic radial equation for such spacetimes has not been explicitly derived. In the following, we derive this equation by generalizing the Teukolsky-Chandrasekhar procedure \cite{Chandrasekhar:1985kt}. This can be used in other contexts. \\

To this aim, we have used the Newmann-Penrose formalism \cite{NewPen}, which is, among other desirable properties, well-suited for spherical BHs. In this formalism, we have chosen a null basis consisting of a pair of real null vectors \textbf{l} and \textbf{n} and a pair of complex conjugate null vectors \textbf{m} and $\overline{\textbf{m}}$:

\be
\textbf{l}.\textbf{l}=\textbf{n}.\textbf{n}=\textbf{m}.\textbf{m}=\overline{\textbf{m}}.\overline{\textbf{m}}=0.
\label{pro1}
\ee

\noindent The orthoganility conditions are imposed: 

\be
\textbf{l}.\textbf{m}=\textbf{l}.\overline{\textbf{m}}=\textbf{n}.\textbf{m}=\textbf{n}.\overline{\textbf{m}}=0.
\label{pro2}
\ee

\noindent We also require the following normalization: 

\be
\textbf{l}.\textbf{n}=1 \ \ \mathrm{and} \ \ \textbf{m}.\overline{\textbf{m}}=-1.
\label{pro3}
\ee

\noindent This normalization condition is not necessary in the Newmann-Penrose formalism, but it is convenient for our purpose. Any basis with the properties given by Eqs (\ref{pro1}), (\ref{pro2}) and (\ref{pro3}) can be used. We choose the basis vectors:

\bea
l^i=\frac{1}{\sqrt{2}} \left( \frac{1}{\sqrt{G}}, -\sqrt{F}, 0, 0 \right), \\
n^i=\frac{1}{\sqrt{2}} \left( \frac{1}{\sqrt{G}}, \sqrt{F}, 0, 0 \right), \\
m^i=\frac{1}{\sqrt{2}} \left( 0, 0, \frac{1}{\sqrt{H}},\frac{i}{\sqrt{H} sin \theta}\right),  \\
\overline{m}^i=\frac{1}{\sqrt{2}} \left( 0, 0, \frac{1}{\sqrt{H}},\frac{-i}{\sqrt{H} sin \theta}\right).
\eea

When $\delta$ tends to zero and $a_0$ vanishes, this basis tends to the Carter tetrad, which can be used to describe a Schwarzschild BH \cite{batic2017}. However, usually, the Kinnersley tetrad is preferred for Schwarzschild BHs \cite{Chandrasekhar:1985kt}. Different choices for the tetrads will lead to different spin coefficients and finally to apparently different, but actually \textit{equivalent}, radial equations. \\

For spin $\frac{1}{2}$ fields,  the wavefunction is represented by a pair of spinors, $P^A$ and $\overline{Q}^{A'}$, with $A=0,1$ and $A'=0,1$.  The Dirac equation in the Newmann-Penrose formalism can be written as \cite{Chandrasekhar:1985kt}: 

\bea
(D + \epsilon - \rho) P^0 + (\delta ^* + \pi - \alpha) P^1  &=& i \mu _* \overline{Q}^{1'}, \label{un}   \\
(\Delta + \mu - \gamma) P^1 + (\delta  + \beta - \tau) P^0 &=& - i \mu _* \overline{Q}^{0'} \label{deux}, \ \ \  \\
(D + \epsilon ^*- \rho ^*) \overline{Q}^{0'} + (\delta  + \pi ^* - \alpha ^*) \overline{Q}^{1'} &=& - i \mu _*P^1,  \label{trois} \ \ \  \\
(\Delta + \mu _* - \gamma ^* ) \overline{Q}^{1'} + (\delta ^* + \beta ^* - \tau ^* )\overline{Q}^{0'} &=& i \mu _*  P^0,   \ \label{quatre} 
\eea

\noindent with

\be 
 D =l^i \partial _i; \ \ \ \   \Delta =n^i \partial _i; \ \ \ \   \delta =m^i \partial _i; \ \   \ \   \delta ^* = \overline{m}^i \partial _i.
 \label{operator}
 \ee
 
\noindent $\mu _*$ is related to the mass of the fermion $m_e$ by $\mu _* \sqrt{2} = m_e$. The spin-coefficients are derived from the rotation coefficients. In the tetrad formalism (for more details, see, {\it e.g.} \cite{Chandrasekhar:1985kt}), the $\lambda$-symbols are defined as:

\bea 
\lambda  _{abc} = e_{b i, j} [ e_a^i e_c^j - e_a^j e_c^i ],
\label{lsymb}
 \eea
 
\noindent the $a$, $b$ et $c$ indices do indicate the vector of the basis, while the $i$ and $j$ indices are the coordinates. The correspondence reads as $\bm{e_1}=\bm{l}$, $\bm{e_2}=\bm{n}$, $\bm{e_3}=\bm{m}$ and $\bm{e_4}=\bm{\overline{m}}$ with $\bm{e^1}=\bm{e_2}$, $\bm{e^2}=\bm{e_1}$, $\bm{e^3}=-\bm{e_4}$ and $\bm{e^4}=-\bm{e_3}$.  For example, $e_{1 2, 3}$ represents the second composant of $\textbf{l}$, derived with respect to $\theta$. The rotation coefficients are defined as:

\bea 
\gamma  _{cab} = e_c^k e_{a k;i} e_b^i.
\label{rotcoef}
 \eea

\noindent Then, from the $\lambda$-symbols, the rotation coefficients are obtained with the relation:

\bea 
\gamma  _{cab} = \frac{1}{2}( \lambda  _{abc}+\lambda  _{cab}- \lambda  _{bca}),
 \eea

The $\lambda$-symbols (\ref{lsymb}) and the rotation coefficients (\ref{rotcoef}) should not be confused with the spin coefficients $\lambda$ and $\gamma$. The spin coefficients are defined with the rotation coefficients (see Appendix B). So first we have calculated the $\lambda$-symbols and then we have deduced the spin coefficients:

\be 
 \kappa = \sigma = \lambda = \nu = \tau = \pi =0,
 \ee
 
 \be 
 \rho = \mu =\frac{\sqrt{F} H'}{2 \sqrt{2} H},
 \ee

  \be 
 \epsilon =  \gamma =- \frac{\sqrt{F} G'}{4  \sqrt{2} G},
 \ee
 
 \be 
 \alpha = -\beta = - \frac{cot \theta}{2 \sqrt{2 H}}.
 \ee

Given the symmetries, the wavefunctions can be written as $\Psi (t, r, \theta, \phi) = R(r) S(\theta)e^{i(\omega t+m' \phi)}$ where, as for scalars, $\omega$ is the frequency and $m'$ is an integer. We use the following ansatz:
 
\bea
P^0 =\frac{e^{i(\omega t+m'\phi)}}{\sqrt{H(r)}(G(r)F(r))^{\tfrac{1}{8}}} R_+(r) S_+(\theta) , \\    
P^1 =\frac{e^{i(\omega t+m'\phi)}}{\sqrt{H(r)}(G(r)F(r))^{\tfrac{1}{8}}} R_-(r) S_-(\theta) ,  \\
  \overline{Q}^{0'}=-\frac{e^{i(\omega t+m'\phi)}}{\sqrt{H(r)}(G(r)F(r))^{\tfrac{1}{8}}} R_+(r) S_-(\theta),  \\    
\overline{Q}^{1'}=\frac{e^{i(\omega t+m'\phi)}}{\sqrt{H(r)}(G(r)F(r))^{\tfrac{1}{8}}} R_-(r) S_+(\theta)  .
\eea
 
This is useful as it makes the system separable into a radial and an angular parts. The normalisation with $1/(\sqrt{H(r)}(G(r)F(r))^{\tfrac{1}{8}})$ is only chosen for convenience. By inserting the previous expressions in Dirac equation (\ref{un}), we obtain:

\be 
- (\sqrt{HF}\mathcal{D}^{\dag} R_+  + i m_e \sqrt{H}R_-) S_+ + R_-\mathcal{L}S_- =0,
\label{unbis}
\ee

\noindent with $\mathcal{D}$ a radial operator

\be 
\mathcal{D}= \partial _r +  \left( \frac{G'}{8G} - \frac{F'}{8F}  \right) + \frac{iw}{\sqrt{GF}},
  \ee
  
\noindent and $\mathcal{L}$ an angular operator
 
 \be 
\mathcal{L}=  \partial _{\theta}+ \frac{m'}{sin \theta} + \frac{cot \theta}{2}. 
\ee

\noindent $\mathcal{D}^{\dag}$ is the complex conjugate of $\mathcal{D}$ and $\mathcal{L}^{\dag}$ is $-\mathcal{L}$ once replacing $\theta$ by $\pi - \theta $. 

Equation \ref{unbis} implies:

\be
\mathcal{L}S_- = \lambda _1 S_+,
\ee

\be
\sqrt{HF}\mathcal{D}^{\dag} R_+  + i m_e \sqrt{H}R_- = \lambda _1 R_-,
\ee

\noindent with $\lambda _1$ a constant of separation. Proceeding in the same way with Eqs. (\ref{deux}), (\ref{trois}) and (\ref{quatre}), three other constants of separation do appear: respectively denoted $\lambda _2$, $\lambda _3$ and $\lambda _4$. Among the eight equations, there is some redundancy and only four are actually independent. The consistency implies: $\lambda _1=\lambda _2=\lambda _3=\lambda _4 \equiv \lambda$. This separation constant $\lambda$ is neither a $\lambda$-symbol nor a spin coefficient, we simply use the  notation of \cite{Chandrasekhar:1985kt}.\\

The Dirac equations finally reduce to the following radial and angular systems:
\be 
\begin{pmatrix}
\sqrt{HF}\mathcal{D} &  - (\lambda + i m_e \sqrt{H}) \\
 -(\lambda - i m_e \sqrt{H})  & \sqrt{HF}\mathcal{D}^{\dag} 
\end{pmatrix}
\begin{pmatrix}
R_-\\
R_+
\end{pmatrix}= 0,
\label{radialsyst}
\ee

\be
\begin{pmatrix}
\mathcal{L} &  - \lambda  \\
\lambda   & \mathcal{L}^{\dag} 
\end{pmatrix}
\begin{pmatrix}
S_- \\
S_+
\end{pmatrix}= 0.
\label{angularsyst}
\ee

By eliminating $R_-$ in Eq. (\ref{radialsyst}), we obtain the radial equation for $R_+$:

\be 
\sqrt{HF}\mathcal{D} \left( \frac{\sqrt{HF}\mathcal{D}^{\dag} }{\lambda - i m_e \sqrt{H}} R_+ \right)  - (\lambda + i m_e \sqrt{H}) R_+ = 0.
\label{radialeq}
\ee

\noindent The radial equation for $R_-$ is the conjugate of Eq. (\ref{radialeq}). This equation generalizes the Teukolsky equation \cite{Teukolsky:1973ha}. The separation constant $\lambda$ is obtained by solving the angular equation, which is the same than in the Schwarzschild case: $\lambda^2=j(j+1)-s(s-1)$ \cite{Kanti}, that is $ \lambda^2=(l+1)^2$ for fermions.

Setting $m_e=0$ leads to:

\be 
\sqrt{HF}\mathcal{D} \left( \sqrt{HF}\mathcal{D}^{\dag}  R_+ \right)  - \lambda ^2 R_+ = 0.
\ee\\

This equation of motion can be used to determine the fermionic cross section. We study the asymptotic solutions, near the horizon and at spacial infinity. The function $R$ is splitted into its real part $U$ and its Imaginary part $V$. Both equations are then solved thanks to Eq. (\ref{radialeq}).\\

\noindent For a massless fermionic field, at the horizon, Eq. (\ref{radialeq}) tends to:

\be
\frac{\partial ^2 R_+}{\partial r^2} + \frac{1}{2(r-r_+)}\frac{\partial R_+}{\partial r} + \left( \frac{\omega ^2}{C_1}+ i \frac{\omega}{\sqrt{C_1}} \right) \frac{R_+}{(r-r_+)^2} =0,
\label{radhor}
 \ee
 
 \noindent with $C_1= \frac{(r_+ - r_-)^2  r_+^4 }{(r_+^4 +a_o^2)^2} $. With respect to the tortoise coordinate $r^*$, Eq. (\ref{radhor}) reads as:
 
 \be
\frac{1}{C_1}\frac{\partial ^2 R_+}{\partial r^{* 2}} - \frac{1}{2 \sqrt{C_1}} \frac{\partial R_+}{\partial r^{* 2}} + \left( \frac{\omega ^2}{C1}+ i \frac{\omega}{\sqrt{C1}} \right) R_+ =0.
\label{radtorhor}
 \ee

The determinant of the characteristic equation of Eq. (\ref{radtorhor}) is $det=\frac{1-16 \omega (4 \omega M + i )}{4 C_1}$. There are two roots but, from the absorption point of view, there should be only an incoming mode at the horizon. The root $x_1$ is therefore chosen with a positive imaginary part. Near the horizon, the radial part reads as:

 \be
 R_+(r^*)= A e^{x_1 r^*},
 \ee

\noindent with $A$ a complex number. As before, we normalize such that  $R_+(r_{ini})=1$, which leads to $\frac{d R_+(r_{ini})}{dr}= \frac{x_1}{\sqrt{G(r_{ini})F(r_{ini})}}$. At spacial infinity, the solution is a plane wave.  \\

It has been shown in \cite{CveticLarsen} that the transmission coefficient for spin $1/2$ fields is given by: 

\be
|A_l |^2= \left| \frac{A^{h}_{in}}{A^{\infty}_{in}} \right| ^2.
\label{coeff}
\ee

\begin{figure}
\centering
    \includegraphics[width=.9\linewidth]{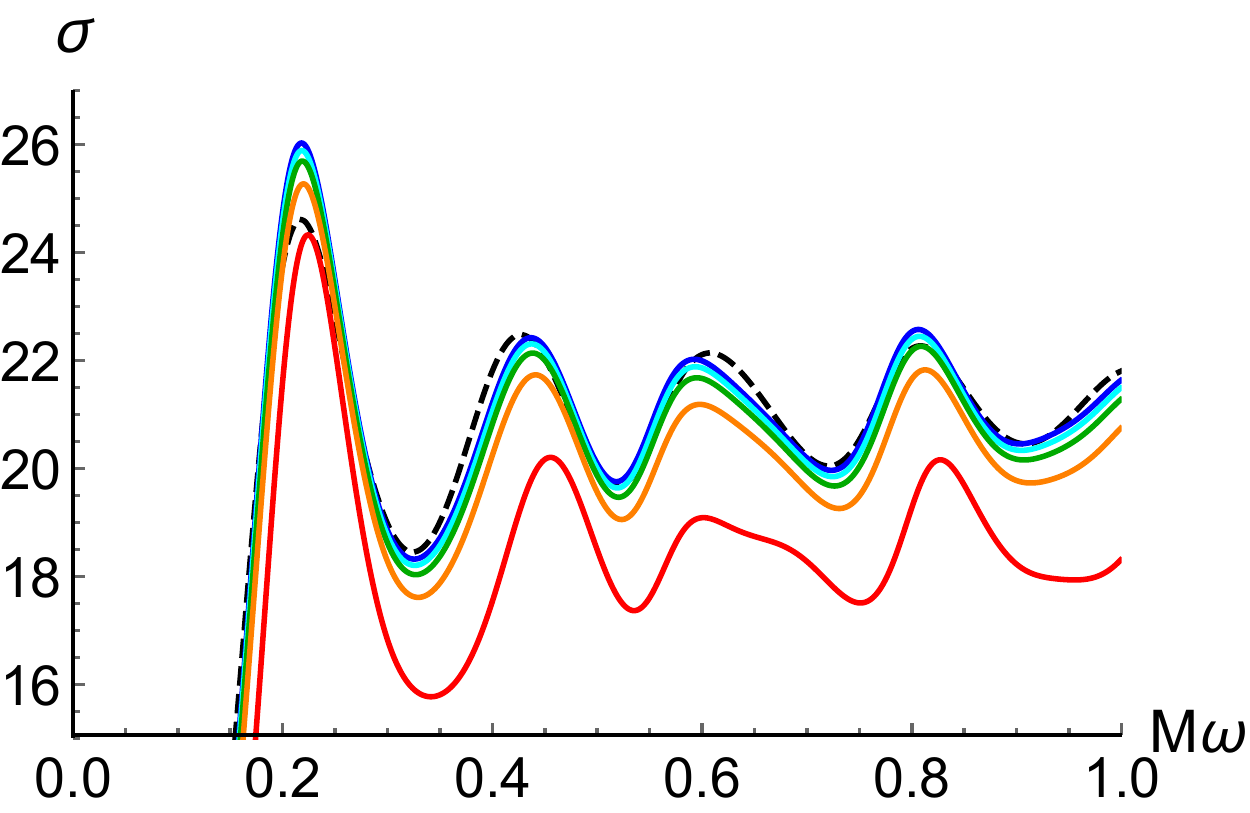}  
 \caption{Emission cross section for a fermionic field, with energy $\omega$, in the background spacetime of a LBH of mass $M$. From bottom to top: $\epsilon=10^{ \{-0.3,-0.6,-0.8,-1,-3\}}$. The dashed dark curve corresponds to the Schwarzschild cross section.}
 \label{res}
 
 \end{figure}

As for the scalar case, we numerically solve Eq. (\ref{radialeq}), fit the solution in order to obtain $A^{\infty}_{in}$ for each $l \leq 10$, and then obtain the cross section. The result is shown in Fig. \ref{res}. Once again, the general trend is to decrease the cross section when the ``quantumness" increases. As the relative effect is getting bigger with an increasing energy of the emitted particle, this should leave a footprint through a distortion of the instantaneous Hawking spectrum which will exhibit slight suppression of its UV tail. \\

 

\begin{figure}
    \includegraphics[width=.9\linewidth]{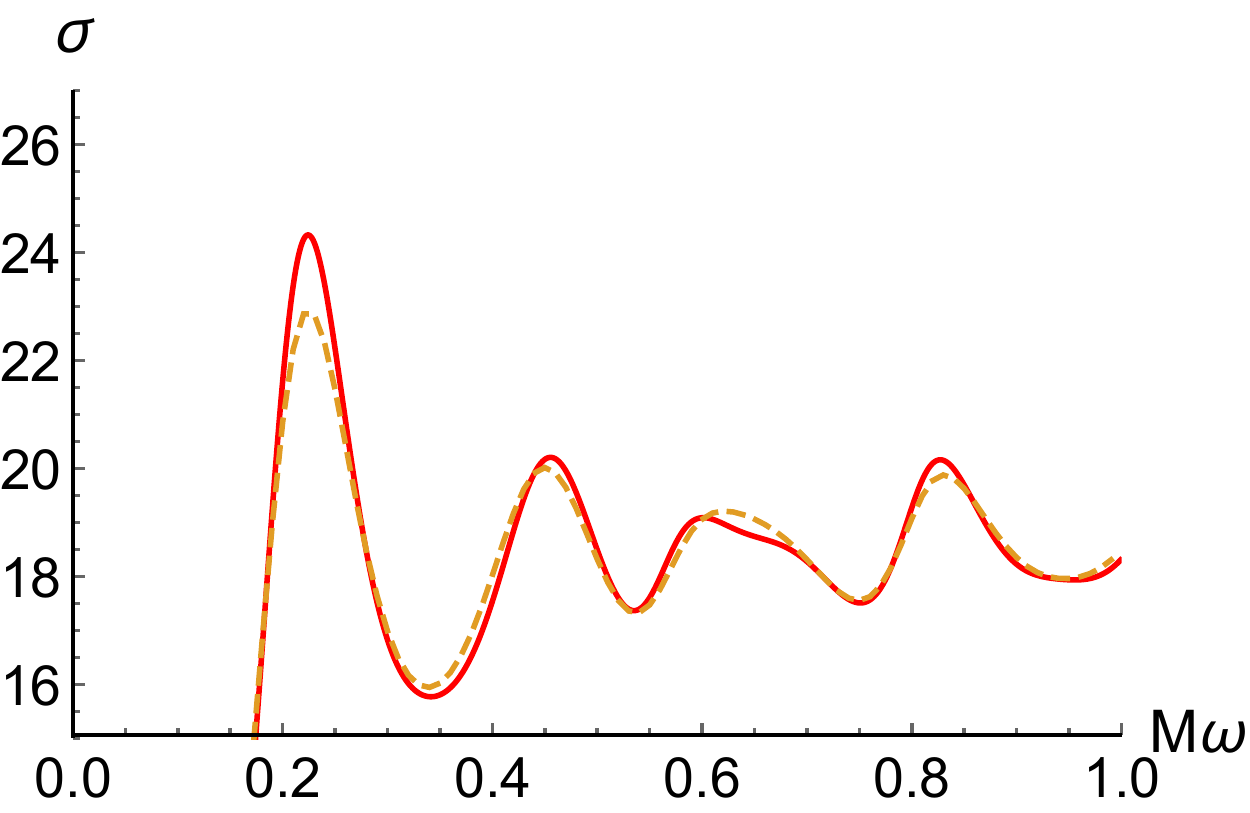}  
 \caption{Emission cross section for a fermionic field, with energy $\omega$, in the background spacetime of a LBH of mass $M$, for  $\epsilon =10^{-0.3}$. The dashed curved corresponds to $a_0=0$ and the plain curve to the usual LQG value, $a_{0}=A_{\text{min}}/8\pi=\sqrt{3}\gamma/2$.}
 \label{a0}
 \end{figure}

Finally, in Fig. \ref{a0}, we show that the effect of sending to 0 the minimum area $a_0$ does not have a dramatic effect. However, choosing a non-vanishing $a_0$ leads to a slight increase of the cross section on the first peak. The cross section itself is of course a continuous function of $a_0$. This parameter has a clearly different influence than the polymerization parameter. 

\section{Conclusion}

In this article, we have studied the propagation of quantum fields in the vicinity of a black hole undergoing quantum gravity corrections. It is shown that the effects are generically small but the trend is quite clear. Phenomenologically, large values of the polymerization parameter could be probed by a decreased cross section, together with a slight frequency shift for fermions.  In addition, the non-vanishing minimum area leaves a specific footprint on the first peak.\\

This sets a framework for futures studies, both in LQG or in modified gravity. In the specific case of loop black holes, it would be most interesting to investigate, using the tools developed in this study, the cross sections for recent BH models published in \cite{BenAchour:2018khr} and \cite{Ashtekar:2018lag,Ashtekar:2018cay}, among others.\\

As the Hawking evaporation of a black hole is considered to be one of the rare possible probes of quantum gravity, it is mandatory to calculate the cross sections for quantum fields in the associated background spacetime. This article is only a first step in this direction for loop quantum gravity. It already shows that different quantum corrections -- still in the LQG framework -- will lead to different effects on the behavior of cross section. This is both useful for accurate calculations of the Hawking spectrum (to refine, {\it e.g.}, what was done in \cite{Barrau:2015ana}) and as a probe, in itself, on the intricate spacetime structure. 

\section{Acknowledgments}

K.M is supported by a grant from the CFM foundation.

\bibliography{refs}

\section{Appendix}

\subsection{Appendix A}

The initial conditions for solving the radial equation (\ref{radialr}) are $R(r_{ini})= A^h_{in}e^{i \omega r^*}=1$ and $\frac{dR(r_{ini})}{dr}=\frac{i \omega}{\sqrt{GF}}$. To solve this complex equation, both the real and the imaginary parts have to be solved. Writing $R(r)=U(r)+ i V(r)$, the initial conditions are:

 \begin{align}
  \begin{aligned}
U(r_{ini})=1, & \\ \frac{dU(r_{ini})}{dr}=0,
  \end{aligned}
  &&
  \begin{aligned}
  V(r_{ini})=0,  &   \\  \frac{dV(r_{ini})}{dr}=\frac{\omega}{\sqrt{GF}}.
   \label{cond}
  \end{aligned}  
 \end{align}

\noindent Far from the BH, we have:
\bea
 U(r)= \frac{a_1}{r} cos(\omega r)+ \frac{b_1}{r} sin(\omega r),  \label{U} \\
 V(r)= \frac{a_2}{r} cos(\omega r)+ \frac{b_2}{r} sin(\omega r),
 \label{V}
\eea
with $a_1=\Re({A^{\infty}_{in}})+\Re({A^{\infty}_{out}})$, $b_1=\Im({A^{\infty}_{out}})-\Im({A^{\infty}_{in}})$, $a_2=\Im({A^{\infty}_{in}})+\Im({A^{\infty}_{out}})$ and $b_2=\Re({A^{\infty}_{in}})-\Re({A^{\infty}_{out}})$. With a fifth order Runge Kutta method, we solve the real and imaginary parts of Eq. (\ref{radialeq}) with the initial conditions given by Eq. (\ref{cond}). At $r_{end}$, we fit the solutions of $U$ and $V$ with functions given in Eqs. (\ref{U}) and (\ref{V}) to obtain the coefficients $a_1$, $b_1$, $a_2$, and $b_2$ so as to deduce $A^{\infty }_{in}$ and $A^{\infty}_{out}$.

\subsection{Appendix B}

The spin coefficients defined with the rotation coefficient are given by:
  \begin{align*}  
 \kappa &=\gamma  _{311} &  \rho &= \gamma  _{314}  & \epsilon &=\frac{1}{2}( \gamma  _{211} +  \gamma  _{341} ) \\  
 \sigma &= \gamma  _{313} & \mu &= \gamma  _{243}  & \gamma &= \frac{1}{2}( \gamma  _{212} +  \gamma  _{342} )  \\
  \lambda &= \gamma  _{244} &  \tau &= \gamma  _{312}  & \alpha &= \frac{1}{2}( \gamma  _{214} +  \gamma  _{344} )  \\
   \nu &= \gamma  _{242} & \pi &= \gamma  _{241}  & \beta &= \frac{1}{2}( \gamma  _{213} +  \gamma  _{343} ) 
  \end{align*}

 \end{document}